\documentclass[aip,apl,reprint,floatfix]{revtex4-1}

\usepackage{graphicx}
\usepackage{comment}
\usepackage{amsmath}
\usepackage{mathptmx}
\usepackage{eucal}
\usepackage{amsfonts}
\usepackage[T1]{fontenc}
\usepackage{hyperref}
\usepackage{lineno}

\begin{document}

\title{Two-Photon Bandwidth of Hyper-Entangled Photons in Complex Media}

\author{Ronen Shekel}
\affiliation{Racah Institute of Physics, The Hebrew University of Jerusalem, Jerusalem, 91904, Israel}
\author{Ohad Lib}
\affiliation{Racah Institute of Physics, The Hebrew University of Jerusalem, Jerusalem, 91904, Israel}
\author{Sébastien M. Popoff}
\affiliation{Institut Langevin, ESPCI Paris, PSL University, CNRS, France}
\author{Yaron Bromberg}
\email[]{Yaron.Bromberg@mail.huji.ac.il}
\affiliation{Racah Institute of Physics, The Hebrew University of Jerusalem, Jerusalem, 91904, Israel}

\date{\today}

\begin{abstract} 
When light propagates through complex media, its output spatial distribution is highly sensitive to its wavelength. This fundamentally limits the bandwidth of applications ranging from imaging to communication. Here, we demonstrate analytically and numerically that the spatial correlations of hyper-entangled photon pairs, simultaneously entangled spatially and spectrally, remain stable across a broad bandwidth: The chromatic modal dispersion experienced by one photon is canceled to first order by its spectrally anti-correlated twin, defining a “two-photon bandwidth” that can far exceed its classical counterpart. We illustrate this modal dispersion cancellation in multimode fibers, thin diffusers and blazed gratings, and demonstrate its utility for broadband wavefront shaping of quantum states. These findings advance our fundamental understanding of quantum light in complex media with applications in quantum imaging, communication, and sensing.
\end{abstract}


\maketitle 

\section{Introduction}
When coherent light propagates through complex media, such as a scattering sample, its spatial output distribution typically depends on the incident wavelength. This dependence can be quantified by the medium's \textit{spectral correlation width}~\cite{Pikalek:19, redding2013all, small2012spatiotemporal, cao2023controlling, dong2025optical}, which describes the wavelength shift required to yield statistically uncorrelated output speckle patterns. A finite spectral correlation width limits the bandwidth available for optical applications, leading to temporal broadening~\cite{dawson1974pulse, tomita1995observation, mccabe2011spatio, small2012spatiotemporal, morales2015delivery} and a reduced speckle contrast~\cite{goodman2007speckle, manni2012versatile, paudel2013focusing}. With classical light, the spectral correlation width can be increased by illuminating the medium with a carefully designed input, known as principal modes, that are immune to chromatic modal dispersion to first order~\cite{fan2005principal, carpenter2015observation}. 

In two-photon experiments, the available bandwidth can be very different from its classical counterpart. When a two-photon state propagates through a complex medium and two-photon correlation measurements are performed, the phases accumulated by both photons must be taken into account~\cite{gerry2023introductory}. Over the past three decades, several aspects of the two-photon bandwidth have been investigated. In the temporal domain, time-energy entanglement has been shown to result in both local ~\cite{steinberg1992dispersion, steinberg1992dispersionPRA} and non-local ~\cite{franson1992nonlocal, franson2009nonclassical, baek2009nonlocal, o2011observations} dispersion cancellation effects, which have been used for optical coherence tomography~\cite{abouraddy2002quantum, nasr2003demonstration}. Similarly, space-momentum entanglement leads to spatial aberration cancellation~\cite{bonato2008even, simon2009spatial, black2019quantum}, an effect leveraged in quantum interferometry ~\cite{simon2009spatial}, imaging~\cite{simon2010odd, filpi2015experimental}, and efficient wavefront shaping~\cite{bajar2025partial, bajar2025rapid}. Additional studies have demonstrated novel phenomena arising from spatially ~\cite{peeters2010observation,abouraddy2012anderson,cande2014transmission,lib2022thermal,lib2022quantum,safadi2023coherent} or spectrally ~\cite{cherroret2011entanglement, cande2013quantum} entangled photons in complex media.

Notably, previous research addressed the effect of either spatial or spectral entanglement. However, the joint effect of spatial and spectral correlations on the available bandwidth in complex media has not been studied. 

In this work, we address this gap by demonstrating analytically and numerically that hyper-entangled photons that are entangled both in their spatial and spectral degrees of freedom experience a novel form of chromatic modal dispersion cancellation. This enables using a significantly larger bandwidth, limited only by the \textit{two-photon spectral correlation width}, which is typically more than an order of magnitude larger than the classical bandwidth. We illustrate our findings initially using multimode fibers, and then extend the results to thin diffusers and blazed gratings. For thin diffusers and blazed gratings we distinguish between two mechanisms that cause wavelength dependence. The first is a geometric scaling effect that produces "exploding speckles" ~\cite{small2012spatiotemporal}, which remains unaffected by the two-photon nature. The second is a "phase wrapping" effect, which is canceled by two-photon interference. Finally, we demonstrate numerically how this cancellation enables broadband wavefront shaping of hyper-entangled photons. Our results enhance the fundamental understanding of quantum light in complex media and open new avenues toward utilizing broadband quantum correlations for advanced photonic technologies.

\section{Results}
When monochromatic light propagates through a complex medium, such as a scattering sample, the resulting output distribution is a high-contrast speckle pattern ~\cite{goodman2007speckle}. This speckle depends on the incident wavelength, and is quantified by the medium's spectral correlation width $\delta\lambda$. When a light source with finite bandwidth $\Delta\lambda$ propagates through such a medium, the output intensity is an incoherent sum of approximately $M\approx\Delta\lambda/\delta\lambda$ independent speckle realizations, resulting in a reduced contrast of $C=1/\sqrt{M}$ ~\cite{goodman2007speckle, paudel2013focusing}. Consequently, the diffusion coefficient of scattering media can be estimated by measuring the contrast of the resulting speckle pattern ~\cite{curry2011direct}. 

\begin{figure*}[htb!]
    \centering
    \includegraphics[width=\linewidth]{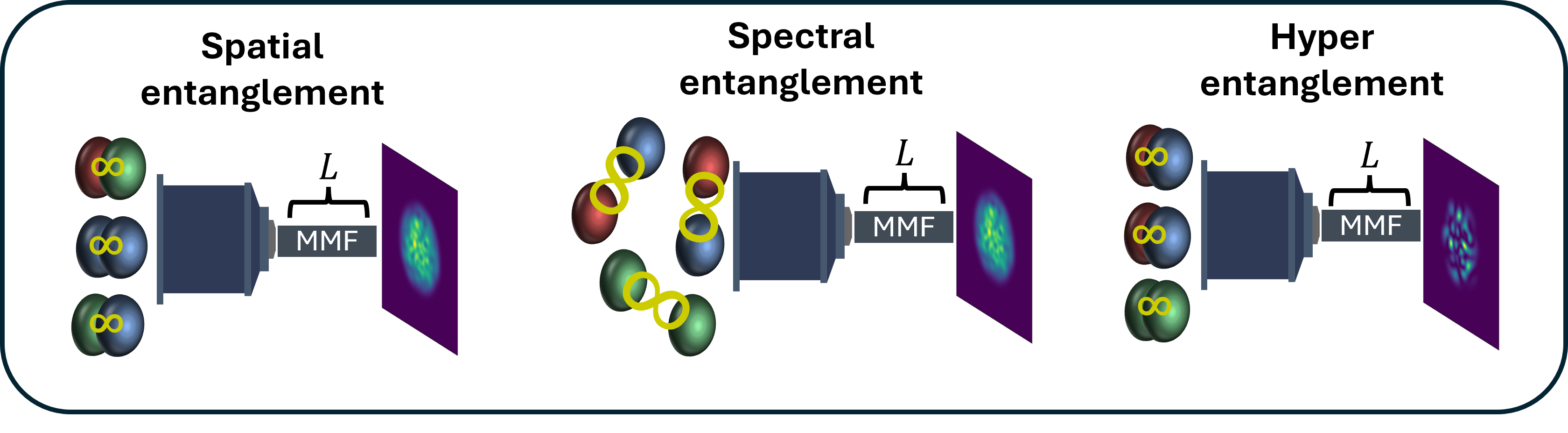} 
    \caption{\label{fig:concept} \textbf{Modal dispersion cancellation for hyper-entangled photons.} Spatially entangled photons occupy correlated spatial modes of the multimode fiber (MMF), but their frequencies are not necessarily correlated. The frequencies of spectrally entangled photons sum to a fixed value, but may occupy different spatial modes. In both cases the resulting two-photon speckle pattern will have a low contrast. Hyper-entangled photons are correlated both spatially and spectrally, resulting in a modal dispersion cancellation effect, manifested by a high contrast two-photon speckle pattern at the fiber output.}
\end{figure*}

We consider the analogous contrast reduction for hyper-entangled photons generated via spontaneous parametric down conversion (SPDC) ~\cite{walborn2010spatial, schneeloch2016introduction}. In SPDC, a strong pump laser beam impinges on a nonlinear crystal, resulting in pairs of entangled photons. From energy conservation, for a monochromatic pump, the sum of the frequencies of both photons must be the frequency of the pump beam, and may be degenerate, with both photons at frequency $\omega_0$, or non-degenerate, such that if one photon has a frequency of $\omega_0+\Delta\omega$, its twin will have a frequency of $\omega_0-\Delta\omega$. In the thin crystal regime ~\cite{walborn2010spatial, schneeloch2016introduction}, with a weakly focused pump, both photons are generated within a small birth-zone, and the spectral-spatial quantum state may be approximated by $\left|\Psi\right\rangle \propto\iint d(\Delta\omega) d\boldsymbol{x}\hat{a}_{\boldsymbol{x},\omega_{0}+\Delta\omega}^{\dagger}\hat{a}_{\boldsymbol{x},\omega_{0}-\Delta\omega}^{\dagger}\left|\text{vac}\right\rangle$, where $\left|\text{vac}\right\rangle$ denotes the vacuum state and $\hat{a}_{\mathbf{x},\omega}^{\dagger}$ is the creation operator in the transverse position $\mathbf{x}$ of a photon with frequency $\omega$. 

The correlations between photon pairs are probed by registering the number of coincidence events, that is, the events in which two photons are detected simultaneously by two detectors. When spatially entangled photons propagate through a scattering medium, and coincidence events are recorded while one detector is kept static and the other scans the transverse plane at the output of the sample, a two-photon speckle is observed ~\cite{peeters2010observation}. Since the SPDC bandwidth often exceeds the spectral correlation width of complex mediums, narrow spectral filters are commonly used to recover high spatial contrast, at the cost of discarding a significant portion of the entangled photon pairs. 

Here we show that for hyper-entangled photons, the bandwidth enabling high-contrast two-photon speckle is not determined by the classical spectral correlation width. Since the phase accumulated by both photons must be taken into account, to first order in $\Delta\omega$ the total phase of degenerate and non-degenerate pairs is equal, and result in the same two-photon speckle, such that high-contrast broadband light may be observed at the output. Intuitively, as depicted in Fig. \ref{fig:concept} the spatial entanglement ensures that both photons experience the same spatial modes of the complex medium, while the spectral entanglement guarantees that the total phase accumulated by degenerate and non-degenerate pairs is equal. 

To observe this modal dispersion cancellation experimentally, one could place different spectral filters at the detectors and compare the contrast of the two-photon speckle when both photons propagate through the complex medium with the contrast obtained when a single heralded photon propagates through the medium, which equals the classical contrast ~\cite{shekel2023pianoq}. From this comparison, the two-photon correlation width can be estimated using the relation $C=1/\sqrt{M}$ ~\cite{goodman2007speckle, paudel2013focusing}.

\subsection{Multimode fibers}
Multimode fibers (MMFs) are an important class of complex media, as they offer a promising platform for increasing the information capacity in both classical ~\cite{puttnam2021space, science2023topological, lu2024empowering} and quantum ~\cite{valencia2020unscrambling, Amitonova2020QKD_MMF, zhou2021high, shekel2023pianoq} communications, as well as for minimally invasive endoscopic imaging applications ~\cite{choi2012scanner, wen2023single, stibuurek2023110, fay2025high}. They have also been utilized for creating programmable quantum circuits ~\cite{leedumrongwatthanakun2020programmable, makowski2024large, goel2024inverse}, and for various other applications ~\cite{cao2023controlling}. However, their finite spectral correlation width ~\cite{redding2013all, Pikalek:19} limits their use with broadband coherent illumination. 

The propagation of classical light through MMFs is described mathematically by decomposing the input field into the fiber’s eigenmodes $f_k(x,y)$ ~\cite{okamoto2021fundamentals}, where $k=1,2,\dots, N$ for a fiber supporting $N$ propagating modes. Ideally, each eigenmode accumulates a mode-specific phase $\exp(i\beta_k L)$ during propagation through a fiber of length $L$, where $\beta_k(\omega)$ denotes the propagation constant which depends on both the frequency and mode number. A coherent superposition of multiple excited modes at the fiber input interferes at the fiber output, creating a frequency-dependent speckle pattern. 

\begin{figure*}[ht!]
    \centering
    \includegraphics[width=\linewidth]{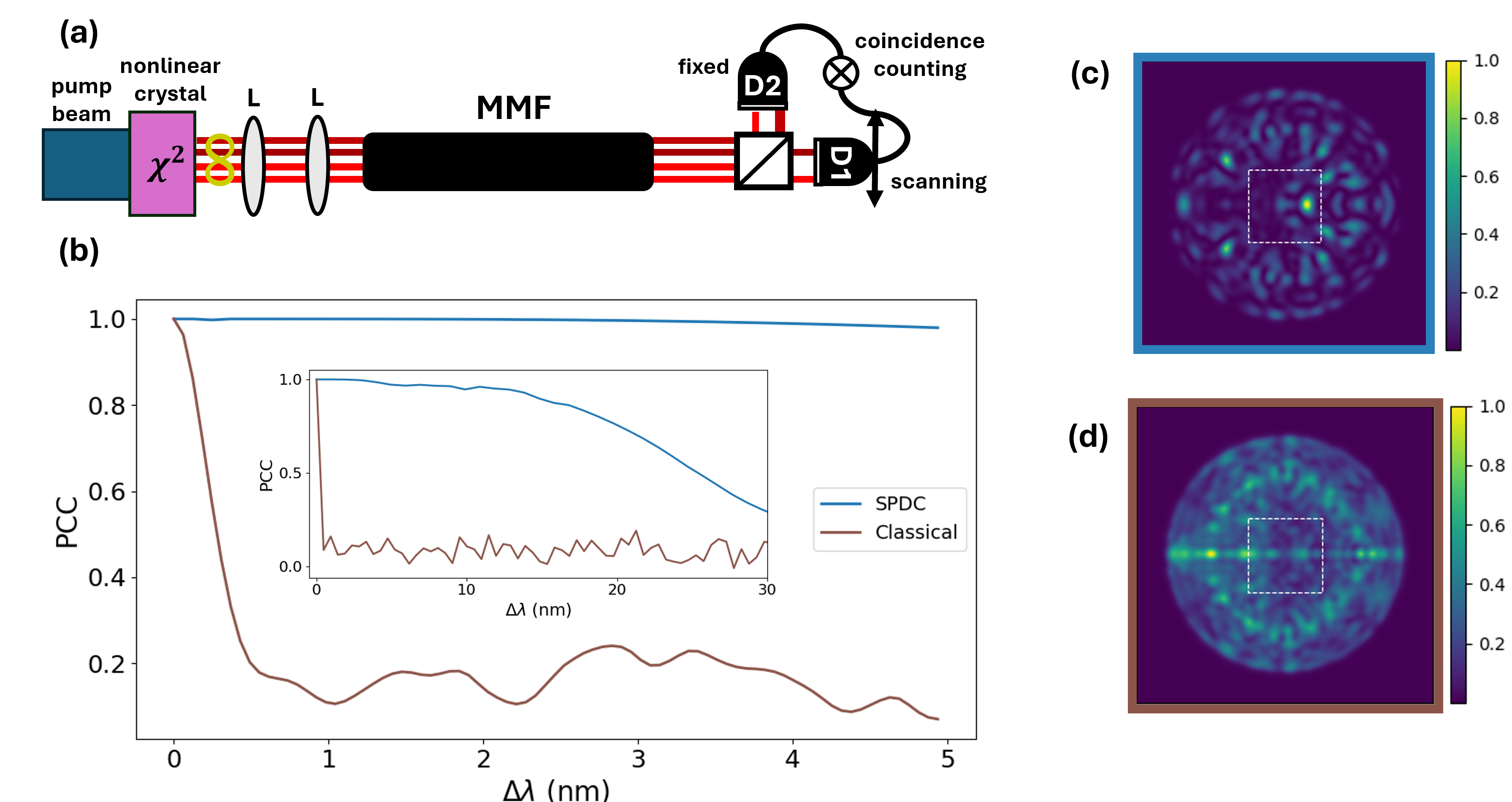} 
    \caption{\label{fig:MMF} \textbf{Modal dispersion cancellation for SPDC photons in a multimode fiber.} (a) A sketch of the proposed setup. Photon pairs, hyper-entangled in the spatial and spectral degrees of freedom, are generated via SPDC and imaged on a multimode fiber (MMF). Coincidence measurements are performed at the distal end of the fiber. L - lens. (b) Numerical results for the Pearson correlation (PCC) between the output speckles of degenerate and non-degenerate photons in the SPDC case (separated by $\Delta\lambda$), and between light with different wavelengths in the classical case. The inset shows a wider wavelength range over which the SPDC patterns begin to decorrelate. (c), (d) The incoherent sums of intensities over all wavelengths up to $\Delta\lambda=5$~nm in the SPDC (c) and classical (d) cases, depicting the expected observation in an experimental scenario. The dashed white squares mark the area over which the Pearson correlations were calculated. The output ring structure visible in (d) is explained by the radial memory effect ~\cite{gokay2025radial}, and the straight line with excess energy stems from the symmetry of the fiber modes $\beta_{\ell,p}=\beta_{-\ell,p}$, where $\ell$ is the orbital angular momentum, as discussed in ~\cite{gokay2025radial}.}
\end{figure*}

As depicted in Fig. \ref{fig:MMF}a, we consider SPDC photons which are imaged onto the input facet of an MMF of length $L$. The general two-photon state in the fiber mode basis can be written as $\left|\Psi\right\rangle =\int d(\Delta\omega) \sum_{n,m}C_{n,m}\hat{a}_{n,\omega_{+}}^{\dagger}\hat{a}_{m,\omega_{-}}^{\dagger}\left|\text{vac}\right\rangle$, where $\hat{a}_{n,\omega}^{\dagger}$ is the creation operator for the fiber mode $n$ with frequency $\omega$. For brevity we defined $\omega_{\pm}\equiv\omega _0\pm\Delta\omega$. However, the spatial entanglement of the state leads to a correlated occupation of the fiber modes, resulting in $\left|\Psi\right\rangle \propto\int d(\Delta\omega) \sum_{n}\hat{a}_{n,\omega_{+}}^{\dagger}\hat{a}_{n,\omega_{-}}^{\dagger}\left|\text{vac}\right\rangle$ (see supplementary material section S1).

For each correlated modal component arising from a term $\hat{a}_{n,\omega_{+}}^{\dagger}\hat{a}_{n,\omega_{-}}^{\dagger}$, the two-photon state of a degenerate pair accumulates a total phase of $2\beta_n(\omega_0)\cdot L$, whereas a non-degenerate pair accumulates a phase of $[\beta_n(\omega_+) + \beta_n(\omega_-)]\cdot L$. The phase difference between the two cases, $\Delta_n^{(\text{2p})}$, determines the stability of the two-photon speckle across a finite bandwidth. Expanding the propagation constant $\beta_n(\omega)$ about $\omega_0$ to second order, we obtain that degenerate and non-degenerate photon pairs accumulate the same phase up to a second order term in $\Delta\omega$:

\begin{equation} \label{eq:MMF_cancelation}
    \begin{split}
    \Delta_n^{(\text{2p})}&=\left[\beta_n\left(\omega_{0}+\Delta\omega\right)+\beta_n\left(\omega_{0}-\Delta\omega\right)-2\beta_n\left(\omega_{0}\right)\right]\cdot L \\ & \approx \frac{\partial^{2}\beta_n}{\partial\omega^{2}}\left(\Delta\omega\right)^{2}\cdot L.
    \end{split}
\end{equation}

When $\Delta_n^{(\text{2p})}-\Delta_m^{(\text{2p})}\ll2\pi ~\forall n,m$, the degenerate and non-degenerate pairs will create similar two-photon speckle patterns, and high contrast spatial correlations should be observed. We thus observe that the \textit{two-photon spectral correlation width} is fundamentally different than the classical spectral correlation width. A full expansion of Eq.~\eqref{eq:MMF_cancelation} to all orders in $\Delta\omega$ will result in a similar cancellation of all the odd-order contributions. 

This cancellation of first-order dispersion is related to previous dispersion cancellation works ~\cite{franson1992nonlocal, steinberg1992dispersion}, but generalizes them to the case of spatially entangled photons propagating through multiple spatial modes. As we discuss below and in the supplementary material section S2, without spatial entanglement the modal occupation will not be completely correlated, and will introduce cross terms $\hat{a}_{n,\omega_{+}}^{\dagger}\hat{a}_{m,\omega_{-}}^{\dagger},~n\neq m$ to the state, for which Eq. \eqref{eq:MMF_cancelation} does not apply. 

We note that the number of modes supported by the fiber $N$ is determined by the fiber core diameter, its numerical aperture, and the optical frequency. In the analytical analysis above, we assumed that the total bandwidth is sufficiently narrow such that the number of modes supported by the MMF and their respective spatial profiles remain effectively constant for all considered wavelengths. In the numerical simulations below we relax this assumption, and calculate the fiber modes for each wavelength separately. 

\subsubsection{Numerical simulations}
To quantify this cancellation of the modal dispersion, we simulate the experiment illustrated in Fig.~\ref{fig:MMF}a. We consider a monochromatic pump stimulating the generation of photon pairs with a central wavelength of $\lambda_0=810$~nm, and assume a thin crystal such that the spatial correlations are perfect. We use the pyMMF solver ~\cite{pymmf}, which models light propagation in MMFs under the scalar approximation, to compute the propagation of these photons through a standard $50~\mu$m step-index MMF of length $10$cm and numerical aperture of $0.2$. As depicted in Fig.~\ref{fig:MMF}b, the Pearson correlation (PCC) between degenerate ($\omega_+=\omega_-$) and non-degenerate ($\omega_+-\omega_-=\Delta\omega$) two-photon speckle patterns remains high for a wide bandwidth. For comparison, we show the Pearson correlation between the intensities of speckle patterns obtained with a classical source at $\omega_0$ and at $\omega_0+\Delta\omega$, which exhibits a much faster decay. All curves are obtained by averaging over 23 different realizations. In the classical case, the realizations differ by the field at the fiber input facet: a tightly focused Gaussian with a waist of $0.6~\mu \text{m}$ whose position is varied relative to the fiber center for each realization. In the SPDC case, the realizations differ by the position of the fixed detector, whose image at the output facet of the fiber corresponds to a Gaussian with a waist of $0.6\mu \text{m}$. The divergence of the Gaussian modes is chosen to be much larger than the numerical aperture of the fiber, ensuring that all fiber modes are both excited and collected. 

Without spectral filtering, measuring the classical intensity or the two-photon correlations result in an incoherent sum of the corresponding speckle patterns at different wavelengths. In Fig.~\ref{fig:MMF}c and Fig. \ref{fig:MMF}d we show these incoherent sums, resulting in a high contrast two-photon speckle in the SPDC case, and a much lower contrast in the classical case. The full simulation code may be found in ~\cite{qdc2025}. 

\subsubsection{Sensitivity to experimental imperfections}
A key assumption in the derivation of modal dispersion cancellation is that the two photons occupy the same fiber modes throughout the propagation in the fibers. Mode mixing in the fiber will therefore degrade the effect. Imperfect imaging between the nonlinear crystal and the input facet of the fiber will also reduce the cancellation.  

It is instructive to illustrate this in Klyshko's advanced wave picture (AWP) ~\cite{klyshko1988simple, belinskii1994two, arruda2018klyshko, shekel2024shaping, zheng2024AWP}. In the AWP, the two-photon correlations can be calculated and analyzed using an analogous classical experiment, in which one of the single-photon detectors is replaced by a classical coherent light source. Light from this fictitious source propagates backwards through the optical system towards the crystal plane. In the AWP the crystal acts as a mirror, which also switches the frequency of the reflected light from $\omega_+$ to $\omega_-$ ~\cite{zheng2024AWP}. The reflected beam returns through the optical setup to the second detector. The AWP guarantees that the resulting intensity pattern in this classical experiment will be identical to the two-photon spatial distribution of the corresponding quantum experiment. 

In the absence of mode mixing in the fiber and perfect imaging between the crystal and the input facet of the fiber, since light propagates through the fiber twice, once with frequency $\omega_+$ and once with frequency $\omega_-$, with the exact same modal distribution, the degenerate and non-degenerate pairs accumulate a similar phase, and one obtains modal dispersion cancellation. 

However, any imperfection in the imaging system can alter the modal distribution between the two effective passes. For example, a defocus aberration is equivalent in the AWP to introducing a segment of free-space propagation between the forward and backward passes through the MMF. This propagation causes diffraction, altering the modal distribution and reducing the cancellation. 


Interestingly, as we show by numerical simulations in the supplementary material section S2, step-index fibers are less sensitive to defocusing compared to graded-index fibers. This observation can be attributed to the same underlying physics as the chromato-axial memory effect ~\cite{vesga2019focusing, devaud2021chromato}: the guided modes of step-index fibers do not mix, to first order, upon free-space propagation. In graded-index fibers, free-space propagation does mix the fiber modes. For these fibers, the sensitivity to defocus aberration is therefore related to the Rayleigh range of a single speckle grain, beyond which mode mixing becomes significant. 


\subsection{Thin diffusers} \label{sec:diffusers}
Thin diffusers represent another class of complex media that are relevant for quantum optics applications. They are often used to model scattering by biological tissue in quantum imaging applications ~\cite{cameron2024adaptive} and by a turbulent atmosphere ~\cite{burger2008simulating, shekel2021shaping} in free-space quantum communications. Scattering and shaping of entangled photons through such diffusers have been studied with both degenerate ~\cite{peeters2010observation, defienne2018adaptive, lib2020real, shekel2024shaping, courme2025non, aarav2025wavefront} and non-degenerate ~\cite{lib2020nondegenerate} SPDC light.

\begin{figure}[b!]
    \centering
    \includegraphics[width=\linewidth]{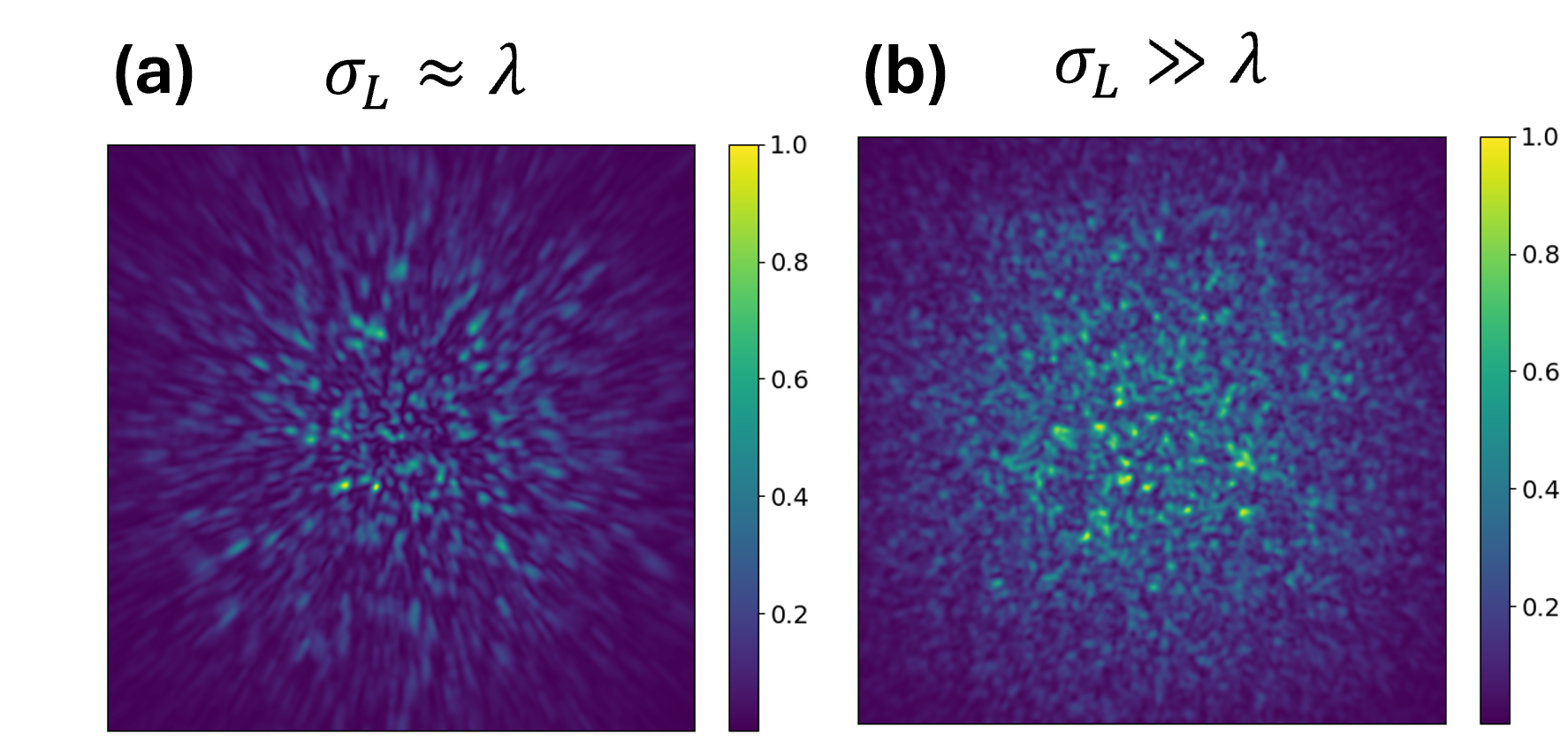} 
    \caption{\label{fig:diffuser1} \textbf{Two types of spatio-temporal effects in thin diffusers.} Simulated far-field intensity distribution of a broadband classical laser ($808\pm40$~nm) scattered by a thin diffuser. (a) Resulting exploding speckles in the regime where the roughness of the diffuser $\sigma_L$ is on the order of the wavelength. The contrast is high close to the optical axis, and is reduced farther away from it due to a geometrical scaling effect. (b) Resulting low-contrast speckle in the regime where the roughness of the diffuser is much larger than the wavelength. The contrast is low everywhere, due to the phase-wrapping effect.}
\end{figure}

As depicted in Fig. \ref{fig:diffuser1}, when broadband classical light propagates through a diffuser and its intensity is measured in the far-field, two distinct mechanisms contribute to a wavelength-dependent output. The first is a geometrical scaling effect, arising from the wavelength-dependent nature of diffraction, that manifests as an "exploding speckle" ~\cite{small2012spatiotemporal}: different wavelengths generate very similar speckle patterns, each scaled by a factor linearly proportional to its wavelength (Fig.~\ref{fig:diffuser1}a). The contrast in this scenario is high close to the optical axis, and is reduced farther away from it. This effect is dominant when the roughness of the diffuser $\sigma_L$ is of the order of the wavelength. The second effect occurs when the diffuser's roughness is much larger than the wavelength, such that the wavelength-dependent phase-wrapping causes different spectral components to experience effectively uncorrelated phase landscapes. In this scenario, different wavelengths produce uncorrelated speckle patterns, resulting in a reduced contrast also on the optical axis, as depicted in Fig.~\ref{fig:diffuser1}b.

\begin{figure*}[tb!]
    \centering
    \includegraphics[width=\linewidth]{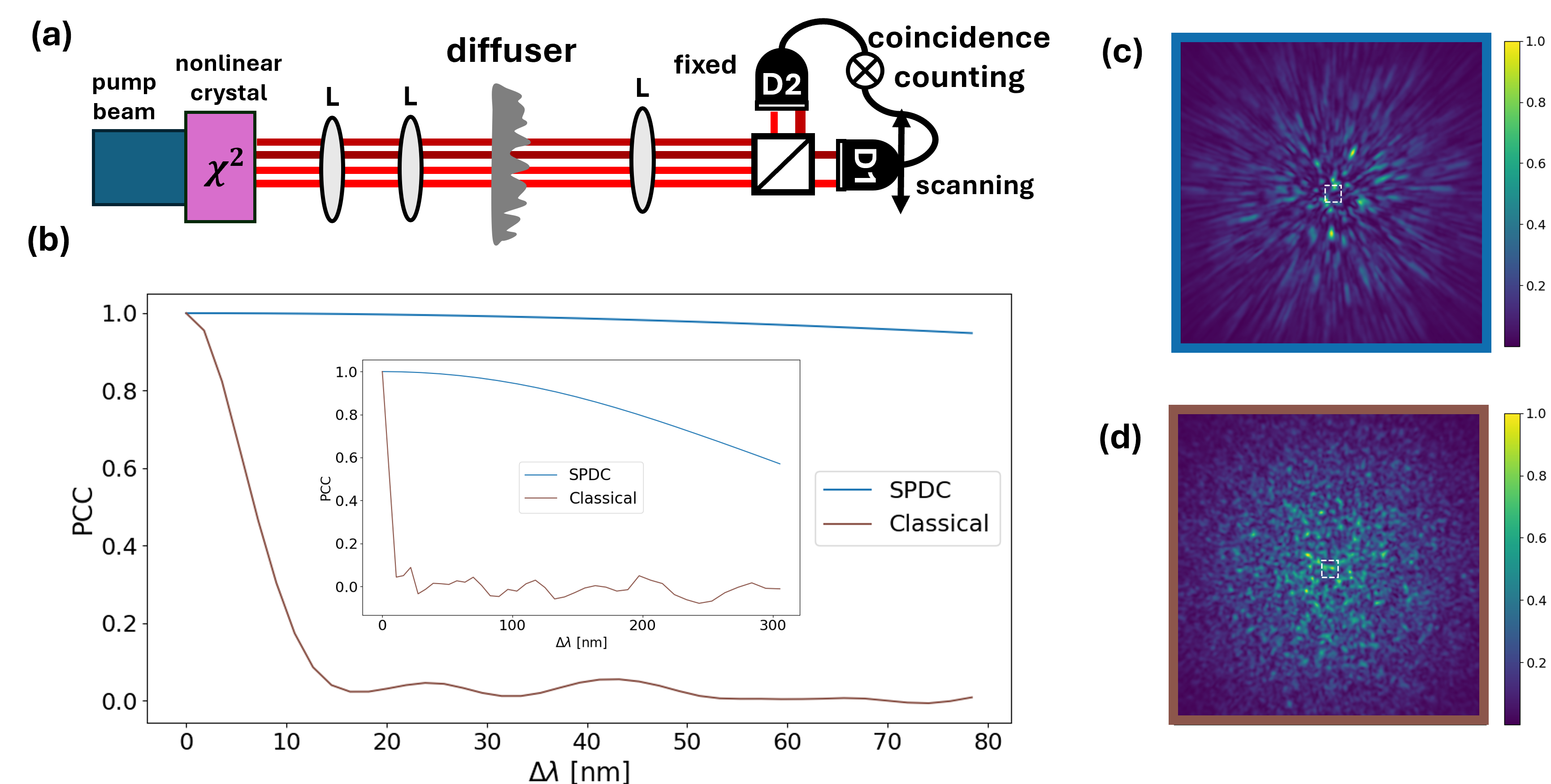} 
    \caption{\label{fig:diffuser2} \textbf{Dispersion cancellation for SPDC photons in thin diffusers.} (a) A sketch of the proposed setup. Spatially and spectrally entangled photons are generated via SPDC, and are imaged to a thin diffuser, after which coincidence measurements are performed in the far-field. L - lens. (b) Numerical results for the Pearson correlation (PCC) between the output speckle patterns of degenerate and non-degenerate photons in the SPDC case, and between light with different wavelengths in the classical case. In both cases, we compare only the region close to the optical axis, as depicted by the dashed white squares. (c), (d) The incoherent sums of intensities over all wavelengths up to $\Delta\lambda=80$~nm in the SPDC (c) and classical (d) cases. For SPDC, the phase-wrapping effect is canceled, and the geometric scaling effect remains, resulting in exploding speckles. See also supplemental movies 1 and 2.}
\end{figure*}

Similar to the discussion for MMFs, when entangled photons propagate through a diffuser, we must take into account the phase accumulated by both photons. As depicted in Fig. \ref{fig:diffuser2}a, we assume the nonlinear crystal is imaged onto the diffuser, such that the spatial entanglement guarantees that both photons pass through the same coherence area on the diffuser. For a diffuser with a refractive index $n$ and with thickness described by $L(x,y)$, the accumulated phase by a photon pair is described by $\phi\left(x,y\right)=(n-1)\left(k_{s}+k_{i}\right)L(x,y)$ where $k_s$ and $k_i$ are the wavenumbers of the signal and idler photons. Since the sum $k_s+k_i$ is conserved, the phase accumulated by a degenerate and non-degenerate pairs is the same, up to the wavelength dependence of the refractive index $n$. The geometric scaling effect, however, is not canceled, and is present irrespective of the diffuser ~\cite{brambila2025certifying}.

Fig.~\ref{fig:diffuser2}b presents our central result for thin diffusers. We first analyze the two-photon case, plotting the Pearson correlation between the speckle patterns of degenerate ($\omega_+=\omega_-$) and non-degenerate ($\omega_+-\omega_-=\Delta\omega$) photon pairs. The correlation remains high across a broad bandwidth. For comparison, the classical correlation, calculated between speckle patterns at $\omega_0$ and $\omega_0+\Delta\omega$, decays much more rapidly. In both cases the correlations were calculated over the region near the optical axis to isolate the phase-wrapping effect and suppress the geometric one, and were averaged over 30 diffuser realizations. The diffuser was modeled using macro-pixels whose thicknesses are uniformly distributed in the range $[0,40\lambda_0]$ for $\lambda_0=808$~nm, and incorporated the material dispersion of a typical polymer diffuser provided to us by RPC Photonics. 

The consequence of this robust quantum correlation is shown in Figs.\ref{fig:diffuser2}c-d. The incoherent sum over the entire SPDC bandwidth results in a high-contrast 'exploding speckle' (Fig. ~\ref{fig:diffuser2}c), while the classical sum yields a low-contrast pattern (Fig. ~\ref{fig:diffuser2}d). The full simulation code, including the polymer dispersion curve used, can be found in ~\cite{qdc2025}.

We note that we assume the diffuser to be much thinner than the Rayleigh range of a Gaussian beam with a waist equal to the diffuser's coherence length at the given wavelength. This ensures that diffraction inside the diffuser is negligible, so the diffuser may be treated as a single phase screen.

\subsubsection{Multi-core fibers as thin diffusers}
Multi-core fibers (MCFs) are emerging as promising platforms for classical applications, such as endoscopic imaging ~\cite{badt2022real, weinberg2024ptychographic} and spatial-division multiplexing (SDM) ~\cite{richardson2013space, melo2025all}, as well as for various high-dimensional quantum applications ~\cite{ding2017high, walborn2021qswitch, ortega2021experimental, achatz2023simultaneous}. Specifically, entangled states of the form $\left|\psi\right\rangle \propto\sum_{i}\hat{a}_{i}^{\dagger}\hat{a}_{i}^{\dagger}\left|\text{vac}\right\rangle$, where $i$ labels the core index, have been demonstrated experimentally in multi-core fibers ~\cite{gomez2021multidimensional}.

Because each core accumulates a slightly different optical phase, set by fabrication imperfections and bend-induced strain, an MCF behaves as a single thin phase screen. This analogy is supported by memory-effect imaging and speckle-correlation studies ~\cite{stasio2015light, porat2016widefield, tsvirkun2016widefield, sivankutty2018single}. Consequently, the phase-wrapping mechanism analyzed for thin diffusers applies unchanged to MCFs, and our theory predicts the same dispersion cancellation effect, underscoring their suitability as controllable platforms for observing quantum interference over broad bandwidths.

\subsection{Blazed gratings}
\begin{figure*}[htbp!]
    \centering
    \includegraphics[width=0.9\linewidth]{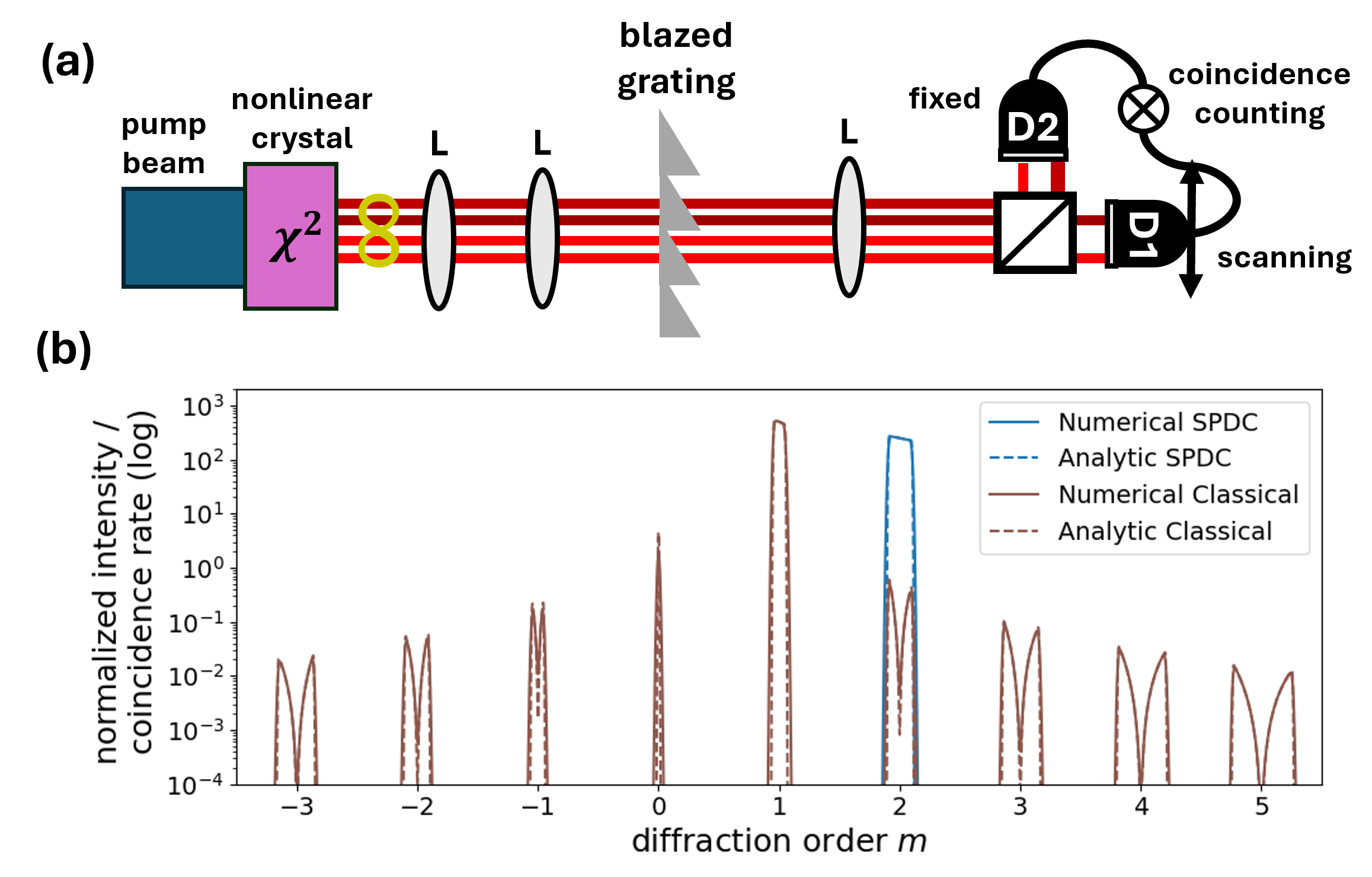} 
    \caption{\label{fig:grating}  \textbf{Dispersion cancellation for SPDC photons in blazed gratings.} (a) A sketch of the proposed setup. Spatially and spectrally entangled photons are generated via SPDC, and are imaged on a blazed grating, after which coincidence measurements are performed, with one of the detectors on the optical axis (order $m=0$), and the other scanning the far field (plotted versus diffraction order $m$). L - lens. (b) The coincidence distribution for SPDC photons, and intensity for classical light at the far-field of the grating. Notably, in the SPDC case all the coincidence events occur in the $m=2$ order, with no leakage to other diffraction orders, due to the cancellation of the phase-wrapping effect. The bandwidth in this simulation is $808\pm40$nm.}
\end{figure*}

Gratings are a textbook example of spatio-temporal optical effects and form the foundation of many spectrometers ~\cite{loewen2018diffraction, jirsa2025fast}: by diffracting different wavelengths into different angles, gratings map frequencies to spatial positions. Blazed gratings are particularly interesting, both because of their higher diffraction efficiency and because they model digital micro-mirror devices ~\cite{popoff2024practical}. As depicted in Fig.~\ref{fig:grating}a, a blazed grating has a periodic triangular phase structure. Mathematically, the optical thickness profile of a grating with period $d$ and design wavelength $\lambda_0$ is described by a sawtooth function:

\begin{equation}
    w\left(x\right)=\lambda_{0}\cdot\frac{\left(x\right)_{\text{mod }d}}{d},
\end{equation}
where $(x)_{\text{mod }d}$ denotes the remainder after division by $d$. At the design wavelength, the accumulated phase ($2\pi w(x)/\lambda_0$) remains continuous and linear across the periods, with no phase jumps. This causes the grating to act as a tilted mirror, directing all incident light into the first diffraction order ($m=+1$).

When broadband classical light passes through such a grating, each wavelength is diffracted at a slightly different angle, creating a spatial rainbow in the $m=1$ diffraction order according to the grating equation. Besides this geometric scaling effect, for wavelengths $\lambda\neq\lambda_0$ the phase jump between grooves is no longer exactly $2\pi$, introducing discontinuities. These discontinuities cause partial leakage into higher diffraction orders, resulting in additional, weaker rainbows. This leakage intensity increases with wavelength mismatch ~\cite{loewen1977grating}, forming a characteristic "double-horn" structure in all diffraction orders except the first, as depicted in the classical intensity distribution in Fig.~\ref{fig:grating}b. 

Next, we consider a two-photon scenario, where one detector is fixed, for instance, at the zeroth diffraction order, and the nonlinear crystal is imaged on the grating. Here, we measure the coincidence rate between the fixed detector and a detector scanning the transverse position in the far-field (in units of diffraction orders). The one-dimensional distributions in Fig. \ref{fig:grating}b are analogous to panels (c) and (d) in Fig. \ref{fig:MMF} and Fig. \ref{fig:diffuser2}. 

Using the AWP, we must consider the photons passing through the grating twice: once at frequency $\omega_0+\Delta\omega$ and once at frequency $\omega_0-\Delta\omega$. Once again, the hyper-entanglement is essential: spatial entanglement ensures both photons traverse the same part of the grating, while spectral entanglement guarantees that the total accumulated phase by the photon pair is independent of $\Delta\omega$. Notably, these two passes combine to produce an effective grating with exactly twice the slope of the classical case. This new effective grating is perfectly blazed independently of $\Delta\omega$. Consequently, for both degenerate and non-degenerate photon pairs, nearly all photons appear in a single diffraction order ~\cite{ostermeyer2009quantum} ($m=+2$ when the static detector is at $m=0$), eliminating the classical phase-wrapping effect. Importantly, while the leakage to other orders is suppressed, the geometric scaling effect remains: within the single occupied diffraction order ($m=2$), the broadband light still forms a rainbow. This is similar to the thin diffuser case where the phase wrapping effect is canceled, while the geometric \textit{exploding speckle} effect remains. The classical and two-photon behaviors are illustrated numerically and analytically in Fig.~\ref{fig:grating}b. A detailed derivation of the analytical model is provided in the supplementary material section S3. The full simulation code can be found in ~\cite{qdc2025}.

\subsection{Wavefront Shaping} \label{sec:WFS}
We have shown that the two-photon speckle pattern from hyper-entangled photons can retain high contrast across bandwidths far exceeding the classical limit. Having a high-contrast pattern is the key prerequisite for wavefront shaping, a powerful technique using spatial light modulators (SLMs) for controlling light in complex media ~\cite{vellekoop2007focusing, gigan2022roadmap}. For classical light, such control is severely limited by the rapid decorrelation of speckle with bandwidth ~\cite{van2011frequency, paudel2013focusing}. Our findings, therefore, suggest that hyper-entangled photons can overcome this limitation, opening new possibilities for shaping broadband quantum correlations for applications such as quantum circuits~\cite{leedumrongwatthanakun2020programmable, makowski2024large, goel2024inverse}.

Interestingly, implementing broadband shaping for MMFs is non-trivial. If the SLM is placed before the fiber, the modal dispersion cancellation effect will be lost, because the phase modulation by the SLM will disrupt the modal correlations, and the state at the fiber input will have terms of the type $\hat{a}_{n,\omega_{+}}^{\dagger}\hat{a}_{m,\omega_{-}}^{\dagger},~n\neq m$. Intuitively, considering the AWP, and as we show explicitly in the supplementary material section S2, the mode population in the first pass through the fiber will be different than in the second pass, since the SLM phases will introduce mode mixing. In order to utilize the high contrast and shape the correlations, the shaping must be performed after the multimode fiber, which has been shown to have fundamentally different features ~\cite{shekel2025fundamental}. 

\begin{figure*}[ht!]
    \centering
    \includegraphics[width=\linewidth]{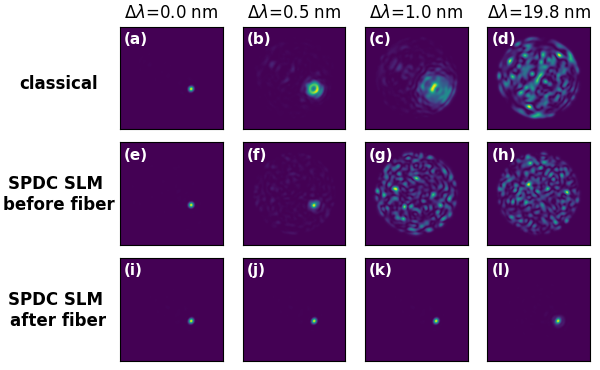} 
    \caption{\label{fig:WFS}  \textbf{Wavefront shaping of broadband SPDC light through an MMF.} ~(a) Classical light focused to a tight spot using an SLM. (b)-(d) The output for the same SLM patterns, as the wavelength is changed. (e) Focused two-photon correlations, induced by an SLM at the input facet of the fiber for a degenerate case. (f)-(h) The output correlations for the non-degenerate case with a difference of $\Delta\lambda$ between the two photons. (i) Focused two-photon correlations, induced by an SLM at the output facet of the fiber, working on one of the photons, for a degenerate case. (j)-(l) The output correlations for the non-degenerate case with a difference of $\Delta\lambda$ between the two photons. See also supplemental movie 3.}
\end{figure*}

In Fig. \ref{fig:WFS} we simulate three different scenarios of performing wavefront shaping using an SLM, with a $20$~cm long step-index fiber: 1) Focusing classical light (Fig. \ref{fig:WFS}a-d). 2) Focusing two-photon correlations, using an SLM at the fiber input (Fig. \ref{fig:WFS}e-h). 3) Focusing two-photon correlations, using an SLM at the fiber output, shaping only one of the photons ~\cite{shekel2025fundamental} (Fig. \ref{fig:WFS}i-l). 

In the classical case, we calculate SLM phases for an initial wavelength, achieving a focal spot (Fig. \ref{fig:WFS}a). We then scan the wavelength while keeping the SLM pattern constant, and observe that the focusing deteriorates within $\approx0.5$~nm (Fig. \ref{fig:WFS}b-d). In the two-photon case with the SLM at the input, we find phases that focus the correlations in the degenerate case at frequency $\omega_0$ (Fig. \ref{fig:WFS}e). Then we scan the frequency difference $\Delta\omega$, and observe that while there is a small partial cancellation effect (see supplementary material section S2), the focus deteriorates within $\approx1$~nm (Fig.~\ref{fig:WFS}f-h). 
In the two-photon case with the SLM at the output, we observe that the same pattern that focuses the correlations in the degenerate case (Fig. \ref{fig:WFS}i), focus also the non-degenerate case up to a bandwidth of $\approx20$~nm where it begins to slowly defocus. 

We note that the SLM at the output, acting as a thin phase screen, has its own finite spectral bandwidth due to the same phase-wrapping effect described in Sec. \ref{sec:diffusers}. In our case this effect was negligible. The full simulation code may be found in ~\cite{qdc2025}. 

Conversely, when the complex medium is a thin diffuser, both the SLM and the diffuser act at the same transverse positions, without mixing between spatial locations. As a result, the SLM may be placed either before or after the diffuser, and the cancellation effect remains intact as long as the SLM is imaged onto the diffuser.

\section{Discussion}
We have shown that hyper-entangled photons, correlated both in space and in frequency, exhibit a modal dispersion cancellation effect when propagating through MMFs, thin diffusers and blazed gratings. The common feature of these systems is that the propagation of light through them is described by phase accumulation without mode mixing: fiber modes for MMFs, and pixel modes for thin elements such as diffusers or blazed gratings. We do not expect this effect to occur in systems with multiple scattering, e.g. white paint. 

In our analysis, we assumed that the photons are perfectly correlated in position, a limiting case corresponding to an infinitely thin crystal and a plane-wave pump beam. In realistic scenarios, however, the dimensions of the crystal and divergence of the pump beam result in a non-zero birth zone and finite emission angles, leading to imperfect spatial correlations ~\cite{walborn2010spatial, schneeloch2016introduction}. In the case of a thin diffuser, this begins to have an effect when the area of the birth zone becomes comparable to the coherence area of the diffuser. Intuitively, in a refined version of the AWP, the crystal acts both as a finite mirror, reflecting the spatial profile of the pump, and as a spatial filter imposed by the phase matching conditions ~\cite{zheng2024AWP}. This spatial filter leads to effective mode mixing, which reduces the cancellation effect. In the limit of a focused pump beam that generates a spatially separable state, this spatial filter transmits a single mode, effectively erasing the mode structure from the first pass in the AWP, and the behavior then reverts to the classical case. In the supplementary material section S2 we show numerically that the cancellation can occur in MMFs for realistic phase-matching parameters. 

Similarly, a finite bandwidth of the SPDC pump beam, which we denote by $\epsilon_p$, results in imperfect spectral correlations between the two photons. If one photon has a frequency of $\omega_0+\Delta\omega$, its twin may have a frequency of $\omega_0-\Delta\omega+\epsilon_p$. This leads to a correction to Eq.~\eqref{eq:MMF_cancelation}, introducing an additional first order term in the pump bandwidth: 

\begin{equation}
    \Delta_{n}^{(\text{2p})} \approx \frac{\partial^{2}\beta_n}{\partial\omega^{2}}\left(\Delta\omega\right)^{2}\cdot L \pm \frac{\partial\beta_n}{\partial\omega}\epsilon_p\cdot L.
\end{equation}

Indeed, only hyper-entangled photons, correlated both spatially and spectrally, result in the chromatic modal dispersion cancellation. 

Finally, to contrast the modal cancellation effect for spectral-spatial hyper-entangled photons we analyzed in this work, we consider an analogous hyper-entangled polarization-spatial scenario. When classical light propagates through a long MMF, strong spatial-polarization mixing occurs, such that the two output polarizations produce uncorrelated speckle patterns. Thus, measuring the output speckle without using a polarizer will result in a reduced contrast of $C=1/\sqrt{2}$. For photons entangled spatially and in polarization, there will be $2^2=4$ two-photon uncorrelated speckle patterns corresponding to the $4$ possible two-photon polarizations. Thus, if no polarizers are introduced before the detectors, the contrast will be reduced to $C=1/\sqrt{4}$, with no cancellation effect. The cancellation occurs in the spectral case since unlike polarization, the spectrum of the photons does not change when propagating through a complex linear system. Thus, for a given detected wavelength of one photon, the wavelength of its pair is known. In the polarization case, on the other hand, all four configurations must be taken into account, since the polarization is spatially mixed during the propagation. 

To summarize, we have shown that when hyper-entangled photons propagate through complex media, phase-wrapping spatio-temporal effects are canceled. We believe our findings provide important insight and deepen the understanding of quantum light in complex media, and open the door for various quantum applications, including broadband wavefront shaping and quantum imaging.

\section{ACKNOWLEDGMENTS}
This research was supported by the Zuckerman STEM Leadership Program, the Israel Science Foundation (grant No. 2497/21). R.S acknowledges the support of the Israeli Council for Higher Education and of the HUJI center for nanoscience and nanotechnology. S.M.P acknowledges the French \textit{Agence Nationale pour la Recherche} grant No. ANR-23-CE42-0010-01 MUFFIN.

\section*{Disclosures}
The authors have no conflicts to disclose.

\section*{DATA AVAILABILITY} 
Code and data underlying the results presented in this paper are available in Ref. ~\cite{qdc2025}.

\section*{SUPPLEMENTARY MATERIAL}
In the attached supplementary material we provide: (1) A derivation of the entangled state in the fiber-mode basis. (2) Extended simulations and analysis of the effect of experimental imperfections, including finite phase-matching, a defocus aberration, and shaping using an SLM on the cancellation effect. (3) A full analytical model for the blazed grating diffraction for both classical and SPDC light.

\bibliography{Main}


\end{document}


\title{Two-Photon Bandwidth of Hyper-Entangled Photons in Complex Media - Supplementary Information}

\author{Ronen Shekel}
\affiliation{Racah Institute of Physics,The Hebrew University of Jerusalem, Jerusalem, 91904, Israel}
\author{Ohad Lib}
\affiliation{Racah Institute of Physics,The Hebrew University of Jerusalem, Jerusalem, 91904, Israel}
\author{Sébastien M. Popoff}
\affiliation{Institut Langevin, ESPCI Paris, PSL University, CNRS, France}
\author{Yaron Bromberg}
\email[]{Yaron.Bromberg@mail.huji.ac.il}
\affiliation{Racah Institute of Physics, The Hebrew University of Jerusalem, Jerusalem, 91904, Israel}

\date{\today}

\pacs{}

\maketitle 

\makeatletter
\renewcommand \thesection{S\@arabic\c@section}
\renewcommand\thetable{S\@arabic\c@table}
\makeatother

\def\thefigure{S\arabic{figure}}
\setcounter{figure}{0}
\renewcommand{\theequation}{S\arabic{equation}}
\setcounter{equation}{0}

\section{Calculation of the state in the fiber-mode basis}\label{sec:state_MMF}
Consider a monochromatic pump beam with a Gaussian angular spectrum of width $\sigma$ and wavelength $\lambda_p$ impinging on a nonlinear crystal of length $L_c$. The biphoton state produced via spontaneous parametric down-conversion (SPDC) can be written as \cite{walborn2010spatial, schneeloch2016introduction}: 

\begin{equation}
    \label{eq:psi_full}
    \left|\Psi_0\right\rangle \;\propto\;
        \iint d(\Delta\omega) d\boldsymbol{q_s}d\boldsymbol{q_i} \exp\!\Bigl[-\tfrac{(\boldsymbol{q_{s}}+\boldsymbol{q_{i}})^{2}}{2\sigma^{2}}\Bigr]\,
        \mathrm{sinc}\!\bigl[\frac{L_c}{4k_{p}}(\boldsymbol{q_{s}}-\boldsymbol{q_{i}})^{2}\bigr]\hat{a}^{\dagger}_{\mathbf{q_s},\omega_0+\Delta\omega}\hat{a}^{\dagger}_{\mathbf{q_i},\omega_0-\Delta\omega}\bigl|\text{vac}\rangle,
\end{equation}
with $k_{p}=2\pi n/\lambda_{p}$ the pump wavenumber in the crystal of refractive index $n$, $\boldsymbol{q_s,q_i}$ the transverse momentum of the signal and idler photons, and $\hat{a}^{\dagger}_{\mathbf{q},\omega}$ the creation operator at transverse momentum $\boldsymbol{q}$ and frequency $\omega$. For brevity we define $\omega_{\pm}\equiv\omega_0\pm \Delta\omega$. 

Under the thin crystal approximation, and considering a well-collimated pump beam, the state may be approximated as an EPR state \cite{howell2004realization}, which can be written in the transverse-momentum and in the transverse-position basis as:

\begin{equation}
    \label{eq:psi_EPR}
    \left|\Psi\right\rangle \;\propto\;
        \iint d(\Delta\omega)~d\boldsymbol{q}\;
        \hat{a}^{\dagger}_{\mathbf{q},\omega_+}\hat{a}^{\dagger}_{\mathbf{-q},\omega_-}\bigl|\text{vac}\rangle
    \;\propto\;
        \iint d(\Delta\omega)~d\boldsymbol{x}\;
        \hat{a}^{\dagger}_{\mathbf{x},\omega_+}\hat{a}^{\dagger}_{\mathbf{x},\omega_-}\bigl|\text{vac}\rangle
\end{equation}
where $\boldsymbol{x}$ is the two-dimensional transverse coordinate.

We express Eq. \eqref{eq:psi_EPR} in the guided-mode basis of a weakly guiding multimode fiber,
whose real, orthonormal mode functions $f_{n}(\boldsymbol{x})$ define the state
\begin{equation}
    \label{eq:Sfq_x}
    |f_{n}\rangle \equiv  \hat{a}_{n}^{\dagger}|\text{vac}\rangle=
        \int d\boldsymbol{x}\;f_{n}(\boldsymbol{x})\hat{a}^{\dagger}_{\mathbf{x}}
        |\text{vac}\rangle,
\end{equation}
where $\hat{a}_{n}^{\dagger}$ is the creation operator of a photon in mode $f_n$. Projecting both photons onto this basis gives the expansion $|\Psi\rangle=\sum_{n,m}C_{n,m}\hat{a}_{n,\omega_{+}}^{\dagger}\hat{a}_{m,\omega_{-}}^{\dagger}|\text{vac}\rangle$, with coefficients

\begin{equation}
    \label{eq:SCkk}
        C_{n,m}=\int f_{n}\left(\boldsymbol{x}\right)f_{m}\left(\boldsymbol{x}\right)d\boldsymbol{x}=\int f_{n}\left(\boldsymbol{x}\right)f_{m}^{\ast}\left(\boldsymbol{x}\right)d\boldsymbol{x}=\delta_{n,m},
\end{equation}
where we used the orthonormality of the fiber guided modes, and their real representation.  Thus the SPDC state is diagonal in the mode basis:
\begin{equation}
    \label{eq:Spsi_modes}
    \left|\Psi\right\rangle \propto\int d(\Delta\omega) \sum_{n}\hat{a}_{n,\omega_{+}}^{\dagger}\hat{a}_{n,\omega_{-}}^{\dagger}\left|\text{vac}\right\rangle.
\end{equation}

In section \ref{SI:sensitivity} below we simulate the case of a finite crystal, where the state is entangled but has weaker correlations, introducing to the state cross terms $\hat{a}_{n,\omega_{+}}^{\dagger}\hat{a}_{m,\omega_{-}}^{\dagger},~n\neq m$, which reduce the cancellation effect. 

We note that this calculation is quite general, and assumes only a lossless waveguide that is invariant in the propagation axes, and which supports  propagating modes. Under the scalar paraxial approximation, the propagating modes of any such waveguide are guaranteed to have a real representation. No geometrical symmetry of the waveguide profile is required.

The same calculation may be carried out in Fourier space using the Fourier transforms of the fiber modes, $\tilde f_n(\boldsymbol{q})$. 
Projecting both photons of Eq.~\eqref{eq:psi_EPR} onto this basis gives
\begin{equation}
    C_{n,m} = \int \tilde f_n(\boldsymbol{q})\, \tilde f_m(-\boldsymbol{q})\, d\boldsymbol{q}.
\end{equation}
Since the spatial mode profiles $f_n(\boldsymbol{x})$ can be chosen to be real, their Fourier transforms satisfy 
$\tilde f_n(-\boldsymbol{q}) = \tilde f_n^{*}(\boldsymbol{q})$, 
and therefore
\begin{equation}
    C_{n,m} = \int \tilde f_n(\boldsymbol{q})\, \tilde f_m^{*}(\boldsymbol{q})\, d\boldsymbol{q}
    = \delta_{n,m}.
\end{equation}

\section{Sensitivity of the cancellation to experimental imperfections} \label{SI:sensitivity}
\subsection{General input state} \label{SI:mode-mixing}
The modal dispersion cancellation effect relies on a perfectly correlated biphoton state $\left|\Psi\right\rangle \propto\int d(\Delta\omega) \sum_{n}\hat{a}_{n,\omega_{+}}^{\dagger}\hat{a}_{n,\omega_{-}}^{\dagger}\left|\text{vac}\right\rangle$ entering the fiber, and on both photons traveling through the fiber with no mode mixing. This, however, may not be the case in a realistic scenario. First, due to finite phase-matching conditions, the state generated via SPDC is not perfectly correlated in space, and hence not perfectly correlated in the modal basis (section \ref{SI:phase_matching}). Second, aberrations in the imaging system between the nonlinear crystal and the fiber (section \ref{SI:defocus}), or applying wavefront shaping with an SLM (section \ref{SI:WFS}) change the state and will introduce non-diagonal terms $\hat{a}_{n,\omega_{+}}^{\dagger}\hat{a}_{m,\omega_{-}}^{\dagger},~n\neq m$. 

Here, we consider a general two-photon state at the fiber input, 

\begin{equation}
\left|\Psi\right\rangle = \int d(\Delta\omega) \sum_{n,m}C_{n,m}\hat{a}_{n,\omega_{+}}^{\dagger}\hat{a}_{m,\omega_{-}}^{\dagger}\left|\text{vac}\right\rangle,    
\end{equation}
propagating through an ideal fiber with no mode mixing. After propagating through a fiber of length $L$, the phase accumulated by a cross term $n\neq m$ is $\left[\beta_n(\omega_0+\Delta\omega)+\beta_m(\omega_0-\Delta\omega)\right]\cdot L$. The phase difference $\Delta_{n,m}^{(\text{cross})}$ for such a cross term between the degenerate and non-degenerate cases is:
\begin{equation}
    \begin{split}
\Delta_{n,m}^{(\text{cross})} 
&= \left[\beta_{n}\!\left(\omega_{0}+\Delta\omega\right)
 + \beta_{m}\!\left(\omega_{0}-\Delta\omega\right)
 - \beta_{n}\!\left(\omega_{0}\right)
 - \beta_{m}\!\left(\omega_{0}\right)\right]\cdot L\\
&\approx 
\left[\frac{\partial\beta_{n}}{\partial\omega}
 - \frac{\partial\beta_{m}}{\partial\omega}\right]\!
\Delta\omega\cdot L
+ \mathcal{O}\!\left((\Delta\omega)^{2}\right).
\end{split}
\end{equation}

When $\Delta_{n,m}^{(\text{cross})}-\Delta_{k,l}^{(\text{cross})}\ll2\pi ~\forall n,m,k,l$, degenerate and non-degenerate pairs will create similar speckle patterns. 

This is different from the classical case. A classical field propagating through an MMF with initial mode occupation $\alpha_n$ is described by $\left|\psi_{\text{cl}}\left(L\right)\right\rangle =\sum_{n}\alpha_{n}e^{i\beta_{n}L}\left|f_{n}\right\rangle$. The phase difference $\Delta_{n}^{(\text{cl})}$ for a specific term between different frequencies is: 

\begin{equation}
    \Delta_{n}^{\left(\text{cl}\right)}=
    \left[\beta_{n}\left(\omega_{0}+\Delta\omega\right)-\beta_{n}\left(\omega_{0}\right)\right]\cdot L
    \approx\frac{\partial\beta_{n}}{\partial\omega}\Delta\omega\cdot L
    +\mathcal{O}\left(\left(\Delta\omega\right)^{2}\right).
\end{equation}

When $\Delta_{n}^{(\text{cl})}-\Delta_{m}^{(\text{cl})}\ll2\pi ~\forall n,m$ the speckle will remain with a high contrast ~\cite{redding2013all}. It is interesting to note the relation $\Delta_{n}^{(\text{cl})}-\Delta_{m}^{(\text{cl})}=\Delta_{n,m}^{(\text{cross})}$.

To summarize, we have identified three distinct decorrelation terms: The classical spectral correlation width is governed by $\Delta_{n}^{\left(\text{cl}\right)}=\frac{\partial\beta_n}{\partial\omega}\Delta\omega\cdot L$. In our two-photon case, for a perfectly correlated state, the limiting factor is determined by the group velocity dispersion (GVD) $\Delta_n^{(\text{2p})}=\frac{\partial^{2}\beta_n}{\partial\omega^{2}}\left(\Delta\omega\right)^{2}\cdot L$. For cross terms, the decorrelation with $\Delta\omega$ is governed by the modal group velocity mismatch (GVM) $\Delta_{n,m}^{(\text{cross})}=\left[\frac{\partial\beta_{n}}{\partial\omega}-\frac{\partial\beta_m}{\partial\omega}\right]\Delta\omega\cdot L$. In all three cases, the spectral correlation width is dictated by the spread of the $\Delta_n$ ($\Delta_{n,m}$) terms over different modes (mode-pairs). In Section II.D of the main text, the effect of these three scales is quantified in the context of wavefront shaping.

\subsection{Effect of finite phase matching} \label{SI:phase_matching}
We now consider the effect of the finite crystal and pump dimensions as described in Eq.~(\ref{eq:psi_full}) above. The mode-basis coefficients could be calculated by 
\begin{equation}
    \label{eq:SCkk_finite}
    C_{n,m}=
        \iint d\boldsymbol{q_{s}}\,d\boldsymbol{q_{i}}\;
        \tilde{f_{n}}(\boldsymbol{q_{s}})\,\tilde{f_{m}}(\boldsymbol{q_{i}})\exp\!\Bigl[-\tfrac{(\boldsymbol{q_{s}}+\boldsymbol{q_{i}})^{2}}{2\sigma^{2}}\Bigr]\,
        \mathrm{sinc}\!\bigl[\frac{L_c}{4k_{p}}(\boldsymbol{q_{s}}-\boldsymbol{q_{i}})^{2}\bigr],
\end{equation}
where $\tilde{f_{k}(}q)$ are the transverse-momentum representations of the guided modes.

\begin{figure}[ht!]
    \centering
    \includegraphics[width=\linewidth]{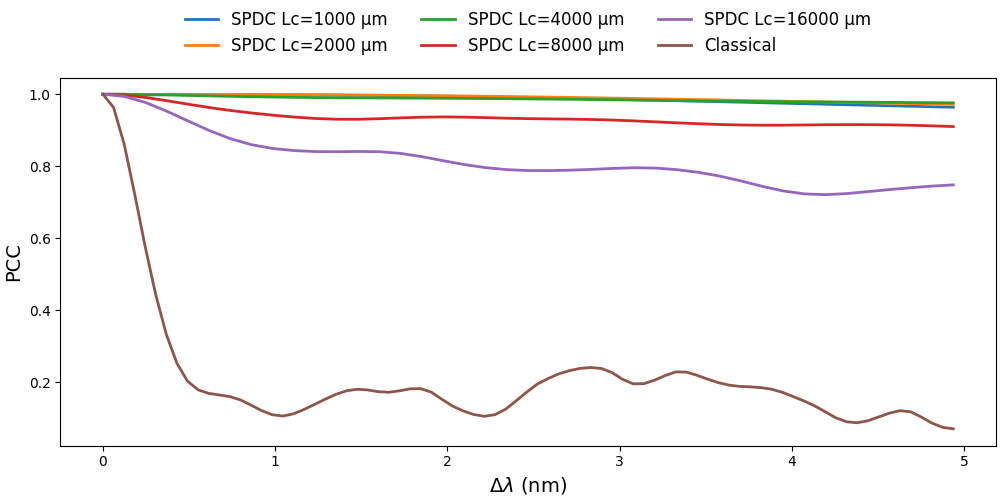} 
    \caption{\label{fig:PCCs_Lcs_SI} \textbf{Finite phase matching: step-index fiber.} In the classical case, the Pearson correlation (PCC) is computed between the output speckle patterns at wavelengths shifted by $\Delta\lambda$. In the SPDC case, the Pearson correlation is between the two-photon speckle patterns of degenerate and non-degenerate pairs. The SPDC results are shown for several crystal thicknesses.}
\end{figure}

\begin{figure}[ht!]
    \centering
    \includegraphics[width=\linewidth]{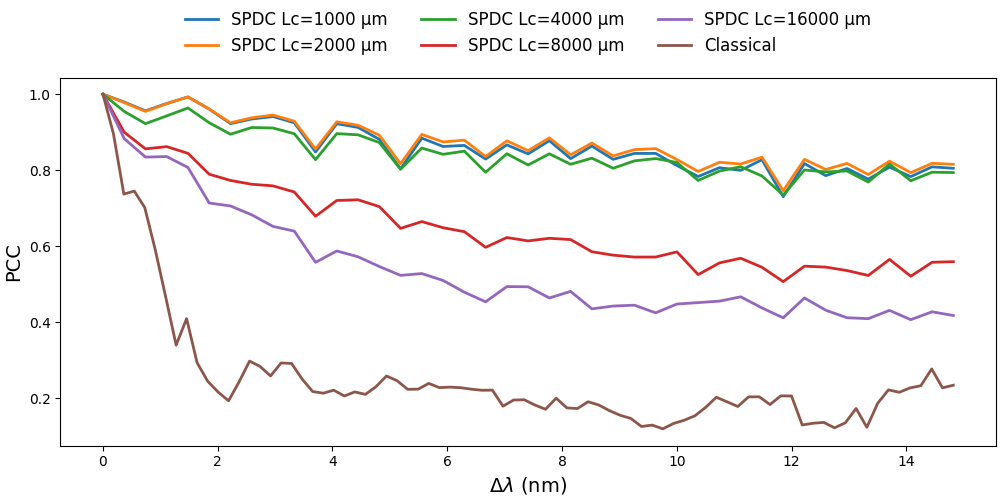} 
    \caption{\label{fig:PCCs_Lcs_GRIN} \textbf{Finite phase matching: graded-index fiber} In the classical case, the Pearson correlation (PCC) is computed between the output speckle patterns at wavelengths shifted by $\Delta\lambda$. In the SPDC case, the Pearson correlation is between the two-photon speckle patterns of degenerate and non-degenerate pairs. The SPDC results are shown for several crystal thicknesses.}
\end{figure}

We simulate the cancellation effect for realistic scenarios with finite phase matching, assuming a realistic pump beam with a waist of $500~\mu m$ and a wavelength of $\lambda_p=405$~nm, and a crystal of lengths $L_c=1,~2,~4,~8,~16$~mm, which are demagnified by a factor of $M=10$ into a multimode fiber, with no defocus aberration. The magnification $M$ stretches the $\boldsymbol{q}$ coordinates, so the width of the sinc function in the fiber plane is $\sqrt{L_c/(4k_{p}\cdot M^2)}$. As depicted in Fig. \ref{fig:PCCs_Lcs_SI} and Fig. \ref{fig:PCCs_Lcs_GRIN}, for step-index and graded-index fibers respectively, the finite phase matching in crystals with lengths up to $4$mm does not suppress the cancellation effect. For thicker crystals the cancellation starts to diminish. The step-index fiber is more immune to the finite phase matching conditions set by the crystal, since it mainly filters out higher order modes, and does not introduce much off-diagonal terms $\hat{a}_{n,\omega_{+}}^{\dagger}\hat{a}_{m,\omega_{-}}^{\dagger},~n\neq m$.

\subsection{Effect of defocus aberration} \label{SI:defocus}
As discussed above, off-diagonal terms $\hat{a}_{n,\omega_{+}}^{\dagger}\hat{a}_{m,\omega_{-}}^{\dagger},~n\neq m$ reduce the cancellation effect. An axial misalignment, or defocus, between the nonlinear crystal's image plane and the fiber facet is a common experimental imperfection that introduces such terms. In the AWP, this defocus is equivalent to a segment of free-space propagation of length $dz$ between the fiber facet and the nonlinear crystal. This propagation causes the beam to diffract, altering the two-photon modal decomposition and thus degrading the cancellation.

Here we quantify this effect, using the same fiber parameters as in the main text ($50 ~\mu$m core diameter, $0.2$ numerical aperture), and with the same averaging procedure, for both step-index and graded-index fibers. The simulation assumes perfect phase matching, and is designed according to the advanced wave picture (AWP): after the first pass through the fiber, free-space propagation is added, followed by a wavelength switch, after which we propagate the photon back to the fiber for a second pass. 

\begin{figure}[ht!]
    \centering
    \includegraphics[width=\linewidth]{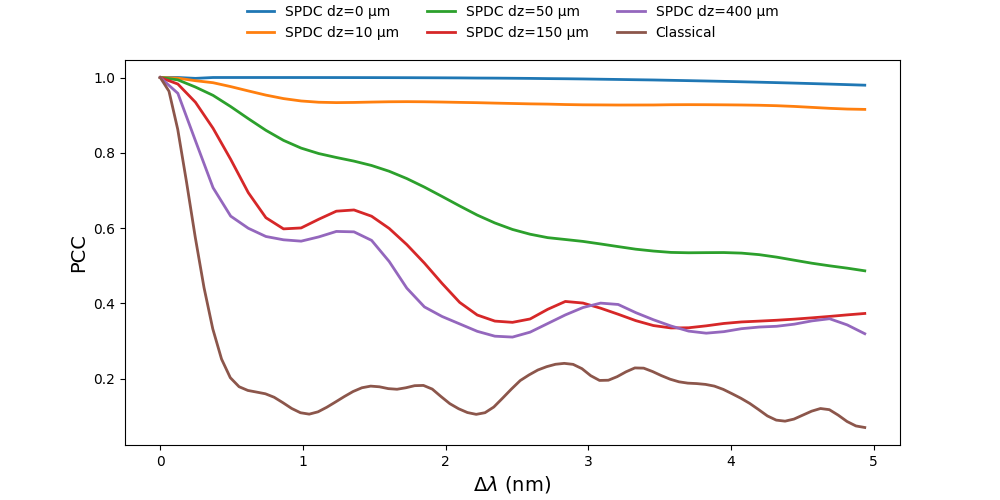} 
    \caption{\label{fig:PCCs_dzs_SI} \textbf{Defocus aberration: step-index fiber.} Numerical results for the Pearson correlation (PCC) between the output speckle patterns of degenerate and non-degenerate photons in the SPDC case, and between light with different wavelengths in the classical case. The results are shown for different values of the defocus distance $dz$ between the nonlinear crystal and the input facet of the MMF.}
\end{figure}

For a step-index fiber (Fig. \ref{fig:PCCs_dzs_SI}), the cancellation is quite robust. This is because, to first order, the guided modes of step-index fibers are not mixed by free-space propagation. Consequently, the cancellation effect degrades significantly only for defocus values of several hundred microns. 

\begin{figure}[ht!]
    \centering
    \includegraphics[width=\linewidth]{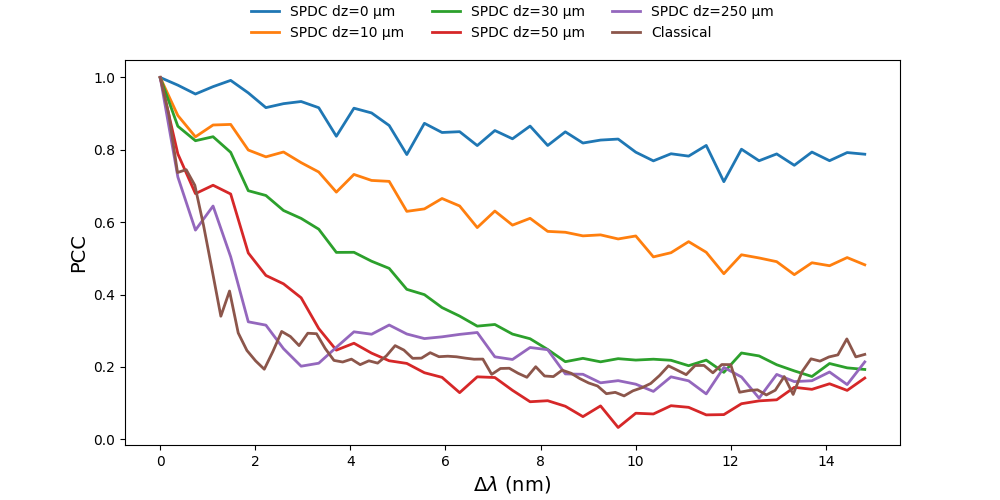} 
    \caption{\label{fig:PCCs_dzs_GRIN} \textbf{Defocus aberration: graded-index fiber.} Numerical results for the Pearson correlation (PCC) between the output speckle patterns of degenerate and non-degenerate photons in the SPDC case, and between light with different wavelengths in the classical case. The results are shown for different values of the defocus distance $dz$ between the nonlinear crystal and the input facet of the MMF.}
\end{figure}

For graded-index fibers (Fig. \ref{fig:PCCs_dzs_GRIN}), free-space propagation does introduce non-diagonal terms $\hat{a}_{n,\omega_{+}}^{\dagger}\hat{a}_{m,\omega_{-}}^{\dagger},~n\neq m$, and the effect is reduced significantly within a few tens of microns. This agrees with the Rayleigh range of a single speckle grain $\lambda/(NA^2)\approx20~\mu$m. The correlation curves are noisier for graded-index fibers, because their propagation constants are grouped into a small number of nearly-degenerate groups, leading to some fluctuations associated with partial revivals that are not fully smoothed out even after averaging over different realizations. 


\subsection{Effect of wavefront shaping before the multimode fiber} \label{SI:WFS}
\begin{figure}[ht!]
    \centering
    \includegraphics[width=\linewidth]{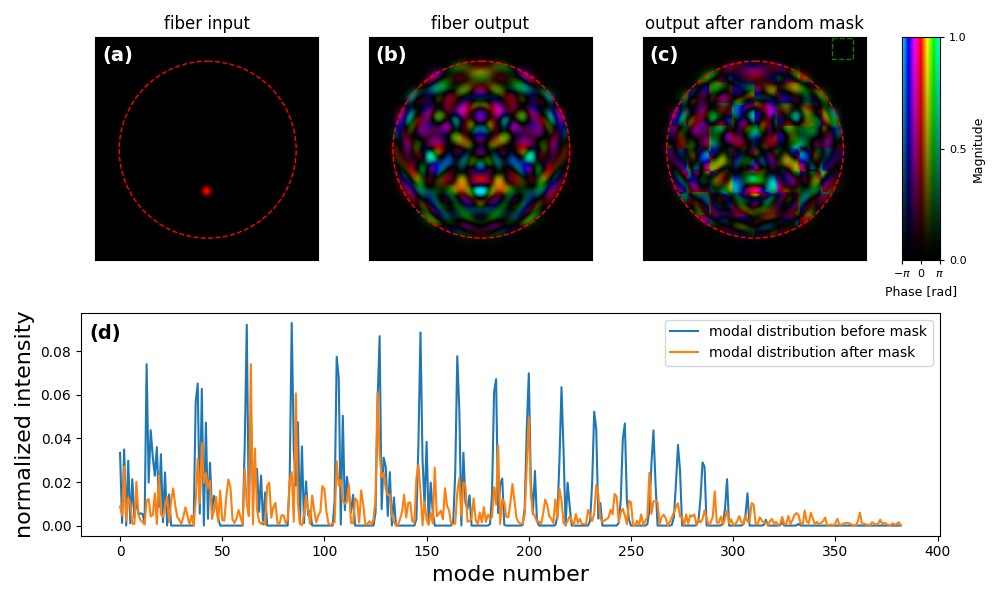} 
    \caption{\label{fig:SLM_mixing_SI} \textbf{SLM inducing mode mixing: step-index fiber} (a) The input field at the fiber facet. (b) The resulting speckle after propagating. (c) the same speckle after passing through a random phase mask, emulating shaping by an SLM. The green dashed square depicts the size of the emulated SLM macro-pixel. (d) The modal distribution before and after passing through the random phase mask. The red dashed circles signify the fiber core.}
\end{figure}

\begin{figure}[ht!]
    \centering
    \includegraphics[width=\linewidth]{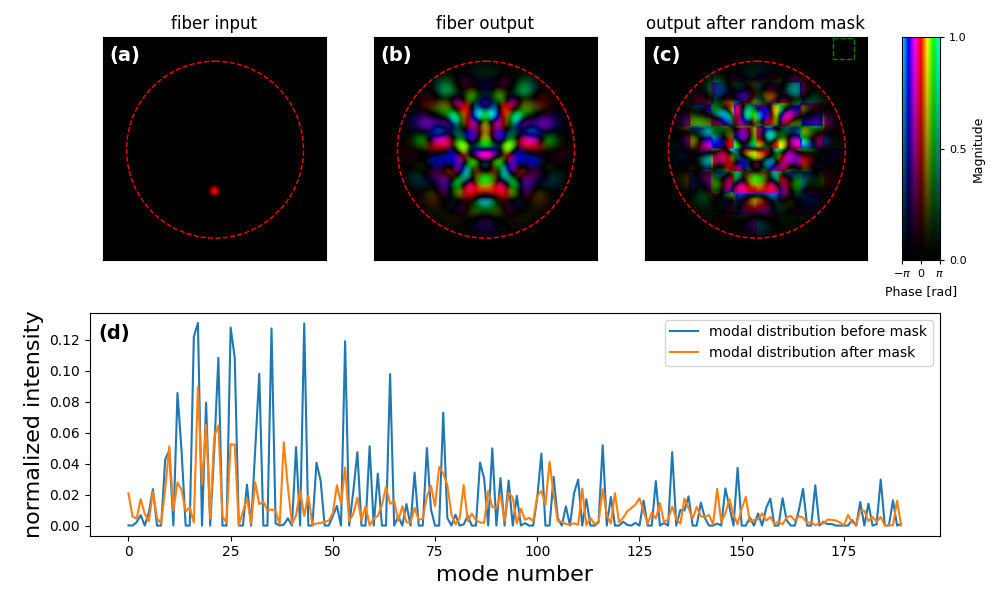} 
    \caption{\label{fig:SLM_mixing_GRIN} \textbf{SLM inducing mode mixing: graded-index fiber} (a) The input field at the fiber facet. (b) The resulting speckle after propagating. (c) The same speckle after passing through a random phase mask, emulating shaping by an SLM. The green dashed square depicts the size of the emulated SLM macro-pixel. (d) The modal distribution before and after passing through the random phase mask. The red dashed circles signify the fiber core.}
\end{figure}

In the main text, we discussed the fact that performing wavefront shaping using an SLM between the crystal and the multimode fiber will result with mode mixing, thereby diminishing the cancellation effect. To quantify this, we simulate for both step-index and graded-index fibers light entering the fiber (Fig. \ref{fig:SLM_mixing_SI}a, Fig. \ref{fig:SLM_mixing_GRIN}a), and propagating through it, resulting in a speckle pattern (Fig. \ref{fig:SLM_mixing_SI}b, Fig. \ref{fig:SLM_mixing_GRIN}b). We then apply random phases to different areas of the field, emulating macro-pixels of an SLM (Fig. \ref{fig:SLM_mixing_SI}c, Fig. \ref{fig:SLM_mixing_GRIN}c). We then compare the modal distribution before and after applying the phase mask, (Fig. \ref{fig:SLM_mixing_SI}d, Fig. \ref{fig:SLM_mixing_GRIN}d), and notice that it changes significantly in both cases. The SLM may gain more control by using a larger number of smaller macro-pixels, but this will simultaneously incur more non-diagonal terms $\hat{a}_{n,\omega_{+}}^{\dagger}\hat{a}_{m,\omega_{-}}^{\dagger},~n\neq m$. 

\section{Analytical model for diffraction from a blazed grating} \label{SI:grating}
To analyze the far-field diffraction pattern from a blazed grating, we perform a Fourier transform from the grating plane to the focal plane of a lens. Under the paraxial approximation, a spatial frequency component $q_x$ in the grating plane maps to position $x' = (f/k) q_x$ at the focal plane, where $f$ is the focal length, $k = 2\pi/\lambda$ is the wavenumber, and $\lambda$ is the wavelength. We define $k_0 = 2\pi/\lambda_0$ as the wavenumber of the design wavelength $\lambda_0$. As depicted in Fig. 5 of the main text, we treat the grating as a transmission phase element, similar to the thin diffuser analysis, and assume negligible material dispersion, i.e. $n(\lambda)$ is constant over the bandwidth. 

\subsection{Classical single-pass analysis}

We model a blazed grating of period $d$ as the convolution of a Dirac comb with a single-groove transfer function. The Dirac comb representing the periodic structure is
\begin{equation}
\mathrm{Comb}(x) = \sum_{n\in\mathbb{Z}}\delta(x-nd),
\end{equation}
whose Fourier transform is:
\begin{equation}
\mathcal{F}\{\mathrm{Comb}\}(q_x) = \frac{2\pi}{d}\sum_{m\in\mathbb{Z}}\delta\left(q_x-\frac{2\pi m}{d}\right).
\end{equation}

Each groove contains a linear phase ramp. For a monochromatic component with wavenumber $k_0 + \delta k$, the transfer function $R(x)$ over one period represents the phase accumulated by the wave:
\begin{equation}
R(x) = \begin{cases}
\exp[i(n-1)(k_0 + \delta k)L(x)], & -d/2 \leq x < d/2 \\
0, & \text{otherwise}
\end{cases},
\end{equation}
where $L(x)=\frac{\lambda_0}{d}x/(n-1)$ is the height profile of the grating, and we assumed a transmission phase grating with index of refraction $n$, similar to the thin diffuser analysis. The blaze angle is related to $d$ through $\tan(\alpha)\cdot(n-1)=\lambda_0/d$ for a first-order blaze configuration (i.e. $m_b=1$). An equivalent phase profile arises in reflection gratings, so the same analysis applies to them as well.

The Fourier transform of this single-groove transfer function is:
\begin{equation}
\mathcal{F}\{R\}(q_x) = d \cdot \mathrm{sinc}\left(\frac{[q_x - (k_0 + \delta k)(\lambda_0/d)]d}{2}\right).
\end{equation}

The infinite grating field is obtained by convolution: $E(x) = [\mathrm{Comb} * R](x)$. Using the convolution theorem, the far-field amplitude of the classical field $E$ is:
\begin{equation} \label{eq:sinc}
\begin{split}
    \widetilde{E}(q_x) & = \mathcal{F}\{\mathrm{Comb}\}(q_x) \cdot \mathcal{F}\{R\}(q_x) \\ & = \frac{2\pi}{d} \sum_m \delta\left(q_x - \frac{2\pi m}{d}\right) \cdot d \cdot \mathrm{sinc}\left(\frac{[2\pi m/d - (k_0 + \delta k)(\lambda_0/d)]d}{2}\right).
    \end{split}
\end{equation}

This gives discrete diffraction orders at $q_x = 2\pi m/d$, each weighted by the corresponding sinc function. 

To account for finite beam size illuminating the grating, we include Gaussian illumination with beam waist $w_0$. This means the incident field is multiplied by $G(x) = \exp[-x^2/w_0^2]$ in the grating plane, with a Fourier transform of $\mathcal{F}\{G\}(q_x) = \sqrt{\pi} w_0 \exp[-q_x^2 w_0^2/4]$. Using the convolution theorem again, the final far-field amplitude becomes a convolution of the discrete diffraction orders with the Gaussian:
\begin{equation}
\widetilde{E}_{\mathrm{final}}(q_x) = \widetilde{E}(q_x) * \mathcal{F}\{G\}(q_x).
\end{equation}

This replaces each delta function at $x'_m = m \lambda f/d$ with a Gaussian peak of width $\sigma_{x'} = \lambda f/(\pi w_0)$. Such a Gaussian appears for each wavelength, and its center is wavelength dependent. 

\subsection{SPDC analysis}

For SPDC photon pairs with signal wavenumber $k_s = k_0 + \delta k$ and idler wavenumber $k_i = k_0 - \delta k$, both photons traverse the grating. The key insight is that the spatial entanglement guarantees that both photons pass through the same location, and their phases add:
\begin{equation} \label{eq:grating_2p}
(n-1)\left[(k_0 + \delta k) + (k_0 - \delta k)\right]\cdot L(x) = (n-1)\cdot2k_0 \cdot L(x).
\end{equation}

This creates an effective phase ramp with a double slope, independent of the frequency detuning $\delta k$. Crucially, since the effective slope is exactly twice the design slope, there are no phase discontinuities (modulo $2\pi$) at the groove boundaries. 

To calculate the coincidence pattern measured with a fixed detector on the optical axis ($m=0$), we utilize Klyshko's advanced wave picture (AWP)~\cite{klyshko1988simple, zheng2024AWP}. In this framework, the coincidence rate is proportional to the intensity of a fictitious classical field that propagates backwards from the fixed detector, reflects from the crystal, and then propagates forward to the scanning detector. 

Propagating the AWP beam from the fixed detector to the grating plane, it then passes through it twice. Propagating to the far-field yields a $\mathrm{sinc}$ expression similar to Eq. \ref{eq:sinc}, but here orders $m\neq2$ vanish: 

\begin{equation}
\mathrm{sinc}\left(\frac{[2\pi m/d - 2k_0(\lambda_0/d)]d}{2}\right) = \mathrm{sinc}\left((m-2)\cdot\pi\right)=0,
\end{equation}

resulting in the simple Fourier transform:
\begin{equation}
\widetilde{E}_{\mathrm{AWP}}(q_x) = \delta(q_x - 4\pi/d)
\end{equation}
corresponding to a single diffraction order at $m = 2$. Assuming the AWP beam covers an area on the grating similar to the classical source, and converting to focal plane coordinates $x' = (f/k) q_x$ we get:
\begin{equation}
\widetilde{E}_{\mathrm{AWP,final}}(x') = \sqrt{\pi} w_0 \frac{k}{f} \exp\left[-\frac{k^2(x' - 2\lambda f/d)^2 w_0^2}{4f^2}\right]
\end{equation}
such that the SPDC peak is centered at $x' = 2\lambda f/d$ with Gaussian width $\sigma_{x'} = \lambda f/(\pi w_0)$.

This calculation demonstrates the key insight of the SPDC configuration: essentially all the coincidence events will be recorded in a single diffraction order, independent of spectral detuning. Within that diffraction order, broadband light still experiences geometric dispersion since each diffraction order's position scales with wavelength as $x'_m = m \lambda f/d$. This analytical model corresponds to the results presented in Fig. 5b of the main text. 

\bibliography{SI}